\documentclass[11pt]{article}
\usepackage{mydef2col}
\usepackage{myArticle}

\usepackage{abstract}
\addto\captionsenglish{}

\title{Floating exercise boundaries for American options in time-inhomogeneous models}
\author{
    \authorstyle{ Andrey Itkin\textsuperscript{1}\,
     and Yerkin Kitapbayev\textsuperscript{2}}
    \newline \newline
    \textsuperscript{1}
    \institution{FRE department, Tandon School of Engineering, New York University, email: \url{aitkin@nyu.edu}} \\
    \textsuperscript{2}
    \institution{ Department of Mathematics, Khalifa University of Science and Technology, email: \url{yerkin.kitapbayev@ku.ac.ae}}
}

\date{\today}

\begin{document}

\maketitle

\lettrineabstract{This paper examines a semi-analytical approach for pricing American options in time-inhomogeneous models characterized by negative interest rates (for equity/FX) or negative convenience yields (for commodities/cryptocurrencies). Under such conditions, exercise boundaries may exhibit a "floating" structure — dynamically appearing and disappearing. For example, a second exercise boundary could emerge within the computational domain and subsequently both could collapse, demanding specialized pricing methodologies.
}

%%%%%%%%%%%%%%%%%%%%%%%%%%%%%%%%%%%%%%%%%%%%%%%%%%%%%%%%%%%%%%%%%%%%%%%%%%%%%%%
\section{Introduction}

Semi-analytical pricing of American options has gained significant attention in recent years, as evidenced in \cite{CarrItkin2020jd, Kitapbayev2021, ItkinMuravey2024jd, Itkin2024jd,ItkinKitapbayev2024} and references therein. The semi-analytical approach can be described as follows: to price an American option (such as a Put option on a stock), we begin with a stochastic model describing the stock's evolution over time through a stochastic differential equation (SDE) with deterministic and perhaps time-inhomogeneous coefficients.

For Markovian stochastic processes, one then can derive a partial differential equation (PDE) that the American option price must solve in the continuation region (where early exercise is suboptimal), subject to specific terminal and boundary conditions. In many cases, this pricing PDE can be transformed into either a Heat or Bessel equation with a general source term and moving boundaries, and solutions of these equations can be obtained analytically using the Generalized Integral Transform (GIT) technique, \cite{ItkinLiptonMuraveyBook}, combined with an extended version of Duhamel's principle, \cite{ItkinMuravey2024jd}. The solution depends explicitly on the exercise boundary, which is not known a priori for American options. However, for every one-factor diffusion and jump-diffusion model examined in the aforementioned literature, the authors derived a non-linear integral Volterra equation of the second kind that determines this exercise boundary. Once solved, the American option price can be represented in an explicit integral form, with the exercise boundary serving as a parameter. Various methods for efficiently solving these equations numerically have been explored.

Note that before these developments, the majority of the relevant papers used the Black-Scholes model with constant coefficients with some exclusions, but also with constant coefficients: the lognormal and CEV diffusion in \cite{KimYu1996}, the American bond options in \cite{Jamshidian1992}, the Heston model in \cite{Chiarella2005}, the Merton jump-diffusion model in \cite{Chiarella2009}, the 3/2 model in \cite{DetempleKitapbayev2017} and the CEV model in \cite{DetempleTian2002}. Our approach presents an attractive alternative to traditional numerical methods for pricing American options, such as, e.g.,  finite-difference (FD) methods. Rather than simultaneously calculating the option price and the exercise boundary (which is defined implicitly), one can explicitly determine the option exercise boundary location. \citep{Andersen2016} advocated this approach for the Black-Scholes model with constant coefficients, proposing an efficient numerical scheme that converges several orders of magnitude faster than conventional FD and tree methods.

As \cite{Itkin2024jd} notes, this methodology offers significant advantages for industrial applications requiring massive computation of American option prices. Indeed, \citep{Andersen2016} demonstrates that various challenges associated with FD and Monte Carlo (MC) approaches often lack uniform and straightforward solutions, necessitating increasingly complex algorithms. The authors emphasize that developing reliable and fast numerical schemes for American option pricing remains an active research area, supporting the view that solving an integral equation for the exercise boundary is more efficient and accurate than numerical PDE solutions.

However, for certain asset classes, American options can exhibit two exercise boundaries, as discussed by \cite{Peskir2007, Battauz2015, AndersenLake2021,Healy2021,Hok2024}. Notably, \cite{AndersenLake2021} demonstrates how to adapt the integral equation technique for the Black-Scholes model with constant coefficients to accommodate this "double boundary" scenario. Such situations can arise when interest rates for FX options are negative, or when both interest rates and convenience yields are negative — conditions that are possible in current market environments. While the principles for constructing integral equations for exercise boundaries can be generalized, the primary challenge lies in properly characterizing the topology of the optimal exercise region.

Another interesting case is discussed in \cite{HilliardNgo2022}, who investigate Bitcoin pricing characteristics and find evidence of jumps and positive convenience yield. Based on their analysis, they conclude that Bitcoin behaves more like a commodity than a currency. However, negative convenience yields are also possible for Bitcoin due to various factors: the threat of unexpected regulation, potential government interference, exposure to hacking risks, and the possibility of lost passwords. Supporting this finding, \cite{Wu2021}  documents negative convenience yields using spot and futures data analyzed through a fractional cointegrated vector autoregressive model, examining data from the Crypto.com Exchange and the Chicago Mercantile Exchange spanning December 18, 2017, to July 31, 2020. Although American options on Bitcoin are not currently traded, their potential future introduction could lead to scenarios where two exercise boundaries emerge as a time-dependent phenomenon.

Obviously, to effectively calibrate any model to market price term-structures, the model must be time-inhomogeneous, with at least deterministic but time-dependent coefficients. This raises an important question: can the GIT technique (or a similar one) for pricing American options under time-inhomogeneous models be extended to address cases with two exercise boundaries? This paper presents an extended approach that addresses this question.

\section{American option price decomposition} \label{S:decomp}

It is known that the price $P(t,x)$ of an American Put option written on $X_s$ at time $s > t \ge 0$ with the strike price $K$ and maturity $T \ge s$ conditional on $X_t = x$ can be determined by solving the optimal stopping problem, \cite{CarrJarrowMyneni1992}
\begin{equation}\label{prob-interest}
P(t,x) = \sup_{t \le \tau \le T} \EQ \left[ D(t,\tau)(K - X_\tau)^+ \right], \qquad D(t,s)=e^{-\int_t^s r(u)du}.
\end{equation}
Here, $D(t,s)$ represents the deterministic discount factor, $r(t)$ is the instantaneous domestic interest rate,
and $\EQ$ denotes the expectation under the risk-neutral measure $\mathbb{Q}$ conditional on $X_t=x$. The supremum is taken over the filtration $\mathcal{F}^X$ which represents all stopping times $\tau$ in the interval $[t,T]$.

Following standard practice, we define the exercise ($\mathcal{E}$) and continuation ($\mathcal{C}$) regions as $\mathcal{E} \bm{\left[ \mathcal{C} \right]} = \Big\{ (u,X_u)\in[0,T)\times (l_x,\infty): \, V(u,X_u) = \bm{\left[ > \right]} K - X_u \Big\}$, where $l_x$ if the left boundary of the $X_t$ domain which in for models considered in this paper could be either $l_x = 0$ or $l_x = -\infty$. These two regions are separated by the early exercise boundaries $X^*_i(t)$, which are time-dependent functions of the time $t$. Here, $i$ can be 1,2 or even 0, potentially representing an empty exercise region.

A key element of our analysis is the decomposition of the American option price into several components. This approach was first proposed by \cite{Kim1990} (also see references in \cite{CarrJarrowMyneni1992}), who decomposed the value of an American put option into two parts: the corresponding European put price and the early exercise premium (EEP). \cite{CarrJarrowMyneni1992} also derived an alternative representation that separates the American put option price into its intrinsic value and time value, demonstrating the equivalence of their results to the McKean equation. While the European option component maintains a consistent form across all cases, the early exercise premium depends on both the optimal exercise boundary (or boundaries) and the structure of the exercise region.

In this paper our focus is on pure diffusion models of the type
\begin{equation} \label{model}
dX_t = \mu(t, X_t) dt + \Sigma(t, X_t) dW_t, \qquad X_t \in \mathbb{R},
\end{equation}
where $W_t$ is the Brownian motion under the risk-neutral measure $\mathbb{Q}$, $\mu(t, X_t)$ is the drift, $\Sigma(t, X_t) > 0$ is the diffusion coefficient. All coefficients of the model are assumed to be continuous and time-inhomogeneous. The \eqref{model} covers a wide class of stochastic processes used in various financial models, including (but not limited to)
\begin{itemize}
\item  The Geometric Brownian motion (GBM) process with deterministic time-dependent parameters with $\mu(t,X_t) = [r(t) - q(t)] X_t, \Sigma(t,X_t) = \sigma(t) X_t$, where $q(t)$ could be a dividend or convenience yield or a foreign interest rate, and $\sigma(t)$ is the log-normal volatility.

\item The Constant Elasticity of Variance (CEV) process with $\mu(t,X_t) = [r(t) - q(t)] X_t, \Sigma(t,X_t) = \sigma(t) X_t^{\beta+1}$, where $|\beta| < 1,  \beta \ne 0$ is a constant.

\item The Ornstein-Uhlenbeck (OU) process with time-dependent parameters and mean-reversion, i.e.,  $\mu(t,X_t) = \kappa(t)(\theta(t) - X_t)$ and $\Sigma(t,X_t) = \sigma(t)$ with $\sigma(t)$ being the normal volatility.

\end{itemize}

To develop a semi-analytical approach to pricing American options under these models, we need generalization of the decomposition of \cite{CarrJarrowMyneni1992} provided for a single exercise boundary case and given by the following Proposition.
\begin{proposition} \label{prop1}
Conditional on $X_t = x$, the American Put price with a {\bf single} exercise boundary $X^*(t)$ can be represented by the following decomposition formula
\begin{align} \label{decompGen}
P \left(t, x \right) &= \EQ \left\{ D(t,T) [K - X_T]^+ \right\} + \int_t^T D(t,u) \EQ \left\{ \left[ r(u)(K-X_u) + \mu(u,X_u) \right] \mathbf{1}_{X_u \in \mathcal{E}} \right\} d u.
\end{align}

\begin{proof}[{\bf Proof}]

Let $X = \left( X_t \right)_{t \geqslant 0}$ be a continuous semimartingale and let $X^*(t): \mathbb{R}_{+} \rightarrow \mathbb{R}$ be a continuous function of bounded variation. Let $F: \mathbb{R}_{+} \times \mathbb{R} \rightarrow \mathbb{R}$ be a $\mathcal{C}^{1,2}$ continuous function on $X_t > X^*(t)$, and $X_t < X^*(t)$, and choose $F(t, X_t) = P(T,X_T)$. Then under some mild regularity conditions using \Ito lemma and change-of-variable formula, \cite{Peskir2005}, the following representation holds
\begin{align} \label{changeVar}
D(t,T) &P(T,X_T) = P(t,x) + \int_t^T D(t,u) \Big[ \mathbb{L}_X P(u,X_u) - r(u) P(u, X_u) \Big] du + M_T + \frac{1}{2} \int_t^T D(t,u) \\
&\times \Big[ P_x(u, X^*(u)+) - P_x(u,X^*(u)-) \Big] d\ell_u(X;X^*), \nonumber \\
\mathbb{L}_X f &:= f_t + \mu(t,x) f_x + \frac{1}{2} \Sigma^2(t,x)f_{xx}, \qquad M_t = \int_0^t D(t,u) P_x(u, X_u) \Sigma(u,X_u) dW_u. \nonumber
\end{align}
Here, $\mathbb{L}_X f$ is an infinitesimal generator of \eqref{model}, $M_t$ is a martingale part of the transformation with $M_0 = 0$, and $\ell(X;X^*)$ is the local time process that $X_t$ spends at the boundary $X^*_t$, see \cite{Peskir2005}. It is important that \eqref{changeVar} holds at the entire domain $(t,x) \in \mathbb{R}_{+} \times \mathbb{R}_{+} \rightarrow \mathbb{R}_{+}$.

Due to the smooth-pasting condition for American options $P_x(t, X^*(t)\pm) = -1$, the integral over the local time in \eqref{changeVar} vanishes. Taking the expectation $\EQ$ of the remaining parts yields
\begin{align} \label{changeVar2}
D(t,T) &\EQ \left[ P(T,X_T)\right] = P(t,x) + \int_t^T D(t,u) \EQ \Big\{ \left[ \mathbb{L}_X P(u,X_u) - r(u) P(u, X_u)\right] \mathbf{1}_{X_u \in \mathcal{C}} \Big\} du \\
&+ \int_t^T D(t,u) \EQ \Big\{ \left[ \mathbb{L}_X P(u,X_u) - r(u) P(u, X_u)\right] \mathbf{1}_{X_u \in \mathcal{E}} \Big\}. \nonumber
\end{align}
The first integral in the right-hand side of this equation vanishes due to the Feynman-Kac theorem valid in the continuation region $X_u \in \mathcal{C}$. For the second integral, in the exercise region $P(u, X_u) = K - X_u$, hence $\mathbb{L}_X P(u,X_u) = - \mu(u, X_u)$. Finally, by rearranging the terms, we obtain \eqref{decompGen}.
\end{proof}
\end{proposition}

\begin{remark}
It is clear that an American put option should not be exercised early when it is out-of-the-money, i.e., when $x \ge K$. To better understand the structure of the optimal exercise region when $x < K$, we can apply Tanaka's formula to the discounted payoff of the Put option (which we aim to maximize) and then take the expected value
\begin{align}
\EQ \left\{ D(t,\tau)(K-X_\tau)^+\right\}=&(K-x)^+ - \EQ \left\{\int_t^\tau D(t,u)H(u,X_u)\mathbf{1}_{X_u<K}du\right\}     + \frac{1}{2} \EQ \left[ \int_t^\tau D(t,u) d\ell_u(X,K) \right], \nonumber
\end{align}
\noindent for $t\in[0,T), \, x > l_x$ and $H(u,X_u)=\mu(u,X_u) + r(u)(K - X_u)$. Since the local time term is monotonically non-decreasing, it follows that exercise is never optimal when $H(t,x) < 0$, implying that $(t,x)\in \mathcal{C}$.
\end{remark}
In \eqref{decompGen}, the first term represents the usual European Put price $P_E\left(t, x \right)$ while the second term is the EEP which depends on the early exercise boundary $X^*(t)$.

In the case of the GBM model with constant coefficients, \eqref{decompGen} was given in \cite{kwok2008mathematical} and reads
\begin{align} \label{decompGenKwok}
P \left(t, x \right) &= \EQ \left\{ D(t,T) [K - X_T]^+ \right\} + \int_t^T D(t,u) \EQ \left\{ \left[ r K - q X_u \right] \mathbf{1}_{X_u < X^*(u)} \right\} d u.
\end{align}
Financially, the last integral represents a sum of discounted cashflows over time produced by exercising the American Put at every moment of time where this is optimal. By the above remark, this is not the case if $H(u, X_u) \equiv r K - q X_u < 0$ or $K < X_u$, hence the exercise region is determined by a system of inequalities: $H(u, X_u) > 0, K \ge X_u$ with $u \to T-$. However, in time-inhomogeneous models, the structure of the exercise region may change at specific points in time as will be seen soon.

Note, that \eqref{decompGenKwok} is also valid for the CEV model and any model which has a drift of the form $(r - q) X_t$ while a particular form  of the volatility function impacts only the transition density.

Using $\psi\left(X_u, u | x, t \right)$ to denote the transition density function of $X_u$ conditional on $X_t = x$, we can rewrite the Put pricing formula \eqref{decompGen} as
\begin{gather} \label{decompDensity}
\resizebox{0.95\textwidth}{!}{$
P \left(t, x \right) = D(t,T) \int_0^K \left(K - X_T\right) \psi\left(X_T, T | x, t \right) d X_T + \int_t^T D(t,u) \int_0^{X^*(u)} H(u, X_u) \psi \left(X_u, u | x, t \right)  d X_u d u.
$}
\end{gather}
Thus, for a given model of the underlying asset, the American Put option price can be explicitly represented as in \eqref{decompDensity} and computed if both the transition density and the exercise boundary are known. When the last integral in \eqref{decompDensity} (the EEP) is positive, the American Put value exceeds its European counterpart. Otherwise, early exercise is never optimal, and the American and European option prices are identical. It is known that for Put options the EEP is positive when either $r > q > 0$ and $X_u < K$, or when $q > r > 0$ and $X_u < K r/q$, resulting in a single exercise boundary. Actually, as was mentioned by an anonymous referee, in this case if $\lim _{u \uparrow T} r(u) > 0$, then the $\operatorname{EEP}\left(\mathrm{t}, \mathrm{X}_{\mathrm{t}}\right)$ is positive for all $X_t$ and $t<T$.

However, as discussed in detail in \cite{AndersenLake2021}, negative values of $r$ and/or $q$ can lead to two exercise boundaries. It can be observed that in this case \eqref{decompGen} of Proposition~\ref{prop1} remains valid, while \eqref{decompDensity} should be replaced by
\begin{equation} \label{decompDensity2}
\resizebox{0.95\textwidth}{!}{$
P \left(t, x \right) = D(t,T) \int_0^K \left(K - X_T\right) \psi\left(X_T, T | x, t \right) d X_T + \int_t^T D(t,u) \int^{X^*(u)}_{X^{**}(u)} H(u, X_u)  \psi\left(X_u, u | x, t \right) d X_u d u,
$}
\end{equation}
\noindent where $0 < X^{**}(u) < X^{*}(u)$, and $X^{*}(u), X^{**}(u)$ are the upper and lower exercise boundaries.

Based on the definition of $H(u, X_u)$ in \eqref{decompDensity}, it is easy to see that the above approach already covers the models with time-dependent coefficients. For instance, for the drift as in \eqref{decompGenKwok}, $H(u, X_u)$ reads
\begin{equation} \label{Htd}
H(u, X_u) = r(u) K - q(u) X_u,
\end{equation}
\noindent while \eqref{decompDensity} maintains its form. This modification, however, has deeper implications: since all model parameters become time-dependent and $H(u, X_u)$ may change sign multiple times. This necessitates careful investigation of the exercise boundaries' topology when evaluating \eqref{decompDensity}.

Using \eqref{decompDensity} with $H(u, X_u)$ given in \eqref{Htd}, one can obtain a nonlinear integral Volterra equation of the second kind for $X^*(t)$ by exploiting the value-matching condition: $P(t,x) = K - X^*(t)$ at $x = X^*(t)$. This yields (note that the same can be done in a general case covered by \eqref{decompGen})
\begin{align} \label{Volterra}
K - X^*(t) &= e^{-r(T-t)} \int_0^K \left(K - X_T\right) \psi\left(X_T, T | x, t \right) d X_T \\
&+ \int_t^T e^{-r(u-t)} \Big[ r(u) K \Psi_1(u, X^*(u) | t, x) - q(u) \Psi_2(u, X^*(u) |t,x) \Big]  \mathbf{1}_{H(u, X_u) > 0}  du, \nonumber \\
\Psi_1(u, X^*(u)  | t.x) &=  \int_0^{X^{*}(u)} \psi\left(X_u, u | x, t \right) d X_u, \qquad
\Psi_2(u, X^*(u)  | t.x) =  \int_0^{X^{*}(u)} X_u \psi\left(X_u, u | x, t \right) d X_u. \nonumber
\end{align}
In \cite{ItkinMuravey2024jd}, the authors provide explicit computations of transition densities for various time-inhomogeneous one-factor models. For both Call and Put options, they also present a detailed derivation of an alternative non-linear integral Volterra equation of the second kind for the exercise boundary. The solutions are obtained analytically by combining the GIT technique with Duhamel's principle. This approach yields an explicit representation of the Green's function (transition density) for the pricing PDE in the continuation region, thus providing both components necessary to evaluate \eqref{decompDensity}.

For the American Call option $C(t,x)$ a representation similar to \eqref{decompDensity} can be derived. For instance, for the GBM model with time-dependent coefficients this can be done by using the Call-Put symmetry for American options, \cite{Kwok2022}. Alternatively, in \cite{CarrJarrowMyneni1992} it is obtained directly to yield
\begin{align} \label{decompDensityCall}
C \left(t, x \right) &= D(t,T) \int_K^\infty \left(X_T - K\right) \psi\left(X_T, T | x, t \right) d X_T \\
&+ \int_t^T D(t,u) \int_{X^*(u)}^\infty H(u, X_u) \psi\left(X_u, u | x, t \right) d X_u d u, \qquad
H(u, X_u) \equiv q X_u - r K. \nonumber
\end{align}
The condition $H(u, X_u) < 0$ should be supplemented by $X_u > K$. However, for a general model in \eqref{model} with time-inhomogeneous coefficients the Call-Put symmetry may not hold, \cite{Detemple2010}.

\section{Models with the drift $\mu(t,X_t) = [r(t) - q(t)] X_t$}

In this section, we examine stochastic processes of the form given in \eqref{model}, where the drift term has the specific structure $\mu(t,X_t) = [r(t) - q(t)] X_t$. This class of processes plays a fundamental role in financial mathematics, as they are extensively used to model various asset classes including equities, foreign exchange rates, commodities, and cryptocurrencies. The widespread application of these processes underscores their theoretical and practical importance.

\subsection{Structure of exercise regions} \label{S:struct}

In \cite{AndersenLake2021} the structure of exercise regions has been extensively analyzed in the context of the Black-Scholes model with constant coefficients, where the underlying asset follows a one-factor GBM stochastic process. Here we extend this analysis by looking at more general and time-inhomogeneous models od the underlying asset through a similar methodology.

Note, that if $q(u) < 0$, $H(u, X_u)$ is an {\it increasing} function of $X_u$, while at $q(u) > 0$ it is a {\it decreasing} function of $X_u$. Further, consider two cases:
\vspace{-1em}
\paragraph{Functions $\bm{ r(u), q(u)}$ do not change sign at $\bm{ u \in [t, T)}$.} From \eqref{decompDensity}, due to convexity of the Put option price, the structure of the exercise regions where the EEP is positive, is determined by two conditions: $H(u, X_u) > 0$ and $X_u < K$ when $u \to T-$. These conditions completely characterize the topology of possible exercise regions, as detailed in Table~\ref{geometry}. When either condition fails to hold, no exercise boundaries exist, and the American Put option price becomes identical to its European counterpart.
\begin{table}[!htb]
\centering
\begin{tabular}{|c|c|c|c|c|}
\toprule
\rowcolor[rgb]{ .792,  .929,  .984}
\multicolumn{1}{|c|}{$\bm q(u)$} &
\multicolumn{1}{c|}{$\bm r(u)$} &
\multicolumn{1}{c|}{$\bm X_u$} &
\multicolumn{1}{c|}{\textbf{Exercise boundaries}} \\ \hline
$\le 0$ &  $q(u) < r(u) < 0$    & $K r(u)/q(u) < X_u < K$   &  two \\ \hline
        &  $r(u) > 0$           & $X_u < K$                 &  one \\ \hline
$\ge 0$ &  $0 < r(u) \le q(u)$  & $X_u < K r(u)/q(u)$       &  one \\ \hline
        &  $r(u) > q(u)$        & $X_u < K$                 &  one \\
\bottomrule
\end{tabular}%
\caption{Geometry of possible exercise regions for the American Put.}
\label{geometry}%
\end{table}%
\vspace{-1.em}
\paragraph{Functions $\bm{ r(u), q(u)}$ do change sign at $\bm{ u \in [t, T)}$.} As shown in \cite{AndersenLake2021}, if an exercise region doesn't exist at time $\tau$, it cannot emerge simply by increasing $\tau$. Therefore, in cases where exercise is never optimal immediately before maturity, it remains non-optimal for all maturities. However, this property may not hold when model parameters are time-dependent.

According to Table~\ref{geometry}, changes in the exercise regime occur at times $\tau_i \in [t,T)$, $i=1,\ldots,N$, where the relationship between $r(u)$ and $q(u)$ differs in the intervals $[\max(t, \tau_{i-1}), \tau_i)$ and $[\tau_i, \min(\tau_{i+1}))$. Such changes occur when either $q(u)$ or $q(u) - r(u)$ changes sign. For example, consider an interval $(\tau_{i-1}, \tau_{i+1})$ where the instantaneous interest rate $r(u)$ maintains its sign. In this case, several scenarios are possible, as illustrated in Table~\ref{signChange}.

Note that Table~\ref{signChange} can be read from either left to right or right to left. This table exhaustively lists all scenarios where $r(u)$, $q(u)$, or $r(u)- q(u)$ may undergo a sign change. Consequently, the EEP in \eqref{decompDensity2} can be expressed as a sum over all such intervals, namely
\begin{align} \label{decompDensitySum}
P \left(t, x \right) &= P_E(t,x) + \sum_{i=0}^{N+1} \int_{\tau_i}^{\tau_{i+1}} D(t,u) \int_{X^{**}(u)}^{X^{*}(u)} H(u, X_u) \psi\left(X_u, u | x, t \right) d X_u d u, \quad \tau_0 = t, \,\, \tau_{N+1} = T,
\end{align}
\noindent where $H(u, X_u) = r(u) K - q(u) X_u$, $X^{**}(u) = 0$ if $N_{EB} = 1$ and $X^{*}(u) = X^{**}(u) = 0$ if $N_{EB} = 0$

\begin{table}[!htb]
\vspace*{-0.5em}
\centering
\scalebox{0.9}{
\begin{tabular}{|c|c|c|c|c|c|c|c|}
\toprule
\rowcolor[rgb]{ .792,  .929,  .984} $\bm{r(u)}$ & $\bm{\tau_{i-1} < t < \tau_{i}}$ & $\bm{X_u}$ & $\bm{N_{EB}}$ & $\bm{q(\tau_i)}$ & $\bm{\tau_{i} < t < \tau_{i+1}}$ & $\bm{X_u}$ & $\bm{N_{EB}}$ \\
\hline
$> 0$   & $q(u) \le 0$ & $X_u < K$ & 1     & 0     & $0 < q(u) < r(u)$ & $X_u  < K$ & 1 \\ \hline
$\ldots$      & $0 \le q(u) < r(u)$ & $X_u  < K$ & 1     & $r(u)$  & $0 < r(u) < q(u)$ & $X_u < K r(u)/q(u)$ & 1 \\ \hline
$< 0$   & $q(u) < r(u) < 0$ &  $K r(u)/q(u) < X_u < K$ & 2     & $r(u)$  & $r(u) < q(u) < 0$ & \text{any}   & 0 \\ \hline
$ = 0$     & $q(u) < 0$ & $X_u < K$ & 1     & 0     & $q(u) > 0$ & \text{any}   & 0 \\
\bottomrule
\end{tabular}%
}
\caption{Switch of exercise regimes when $r(u)$ maintains constant sign over the interval $(\tau_{i-1}, \tau_{i+1})$, (here $N_{EB}$ denotes the number of the exercise boundaries).}
\label{signChange}%
\end{table}%

Also, as discussed in \cite{AndersenLake2021}, in case of two exercise boundaries under some conditions they could collapse into a single point at time $t = t^*$, so for $t < t^*$ no exercise boundaries exist. For the GBM with constant coefficients they define $\sigma^* = |\sqrt{-2 r} - \sqrt{-2 q}|$, and prove that for the American perpetual options with double exercise boundaries (i.e, when $r < q <0$ for calls and $q < r < 0$ for puts), the following statement holds: if $\sigma > \sigma^*$, the boundaries $X^*(t)$ and $X^{**}(t)$ intersect at a finite $t^*$. In other words, they conclude that sufficiently high volatility $\sigma > \sigma^*$ will result in the exercise interval being pinched off at some finite value $t^*$, wherefore it is never optimal to exercise an option at $ t < t^*$. However, no closed-form expression for $t^*$ is known even for the GBM model with constant parameters, although $t^*$ should increase when $\sigma$ increases.

Thus, in time-inhomogeneous models, various scenarios are possible where, for example, a single exercise boundary can disappear or split into two boundaries. This significantly complicates pricing algorithms, especially those where the exercise boundary must be found implicitly together with the American option price, as is inherent to the FD and MC methods (especially, when computing option Greeks). In contrast, the approach that relies on first solving integral equation(s) for the exercise boundary(ies) and then using the semi-analytical representation of the option price as in \eqref{decompDensitySum} makes this problem tractable. It is important to note that the transition density of the problem (the Green's function) does not depend on the number of exercise boundaries or their location, and hence can be found beforehand, as shown in \cite{ItkinMuravey2024jd}.

\begin{remark} \label{remark2}
It is important that the points $\tau_i$, where $r(u)$, $q(u)$, or $r(u) - q(u)$ undergo sign changes, mark transitions in the exercise structure (region). However, these points do not necessarily coincide with the immediate appearance or disappearance of exercise boundaries. This is because the EEP is a time integral over time rather than a single point in time. While a structural change occurs at $t = \tau_i$, the existing boundary has a finite value at this point. Consequently, the actual appearance (from zero) or disappearance of a boundary occurs over a finite time interval, with the duration dependent on the model volatility $\sigma(t)$ (or independent on it in some cases).
\end{remark}

In the following sections, we present examples demonstrating this behavior of American exercise regions in some popular one-factor time-inhomogeneous models.

\subsection{Time-inhomogeneous GBM model} \label{S:model}

Let us consider a one-factor GBM stochastic process with time-dependent coefficients. The stochastic differential equation (SDE) for the underlying asset price $X_t$ is given by:
\begin{equation}
dX_t = (r(t)-q(t)) X_t dt + \sigma(t) X_t dW_t,
\end{equation}
\noindent where $\sigma(t) > 0$ is the volatility, and $W_t$ is a standard Brownian motion under the risk-neutral measure $\mathbb{Q}$. Here, the functions $r(t)$ and $q(t)$ represent the risk-free interest rate and dividend yield respectively, both of which may take negative values.

For this model, the transition density (the Green's function) is well-known, \cite{kwok2008mathematical}. Therefore, the general decomposition formula of the previous section reduces to that proposed by \cite{CarrJarrowMyneni1992} with the modification that all model parameters are time-dependent. The European option component is given by
\begin{align}
P_E(t,x) &= K D(t,T) \Phi\left( d_-(x,K,t,T) \right) - x D_q(t,T) \Phi \left( d_+(x,K,t,T) \right), \qquad  D_q(t,u) = e^{-\int_{t}^u q(s)ds }, \nonumber \\
d_{\pm}(x,y,t,u) &= \frac{1}{\bar{\sigma}\sqrt{u-t}} \left[ \log\frac{y}{x} - \int_t^ u\left(r(s) - q(s) \pm \frac{1}{2} \sigma^2(s) \right) ds \right], \qquad \bar{\sigma} = \left(\frac{1}{u-t}\int_t^u \sigma^2(s)ds\right)^{1/2},
\end{align}
\noindent where $t < u \le T$ and $x, y > 0$.

\subsubsection{Single exercise boundary} \label{caseRpos}

\paragraph{Case $\bm{ r(t)> 0 \, \forall t\in[0,T]}$.}  This is the standard situation when the interest rate $r(t)$ is positive for all $t$. As in the classical case of constant parameters, we have a single  exercise boundary $0\le X^*(t) \le K$ such that $\mathcal{E} = \{(u,X_u)\in[0,T] \times (0,X^*(t))\}$. In other words, the exercise region $\mathcal{E}$ is down-connected.

\begin{figure}[!htb]
\vspace*{-1em}
\centering
\subfloat[]{ \includegraphics[width=0.52\textwidth]{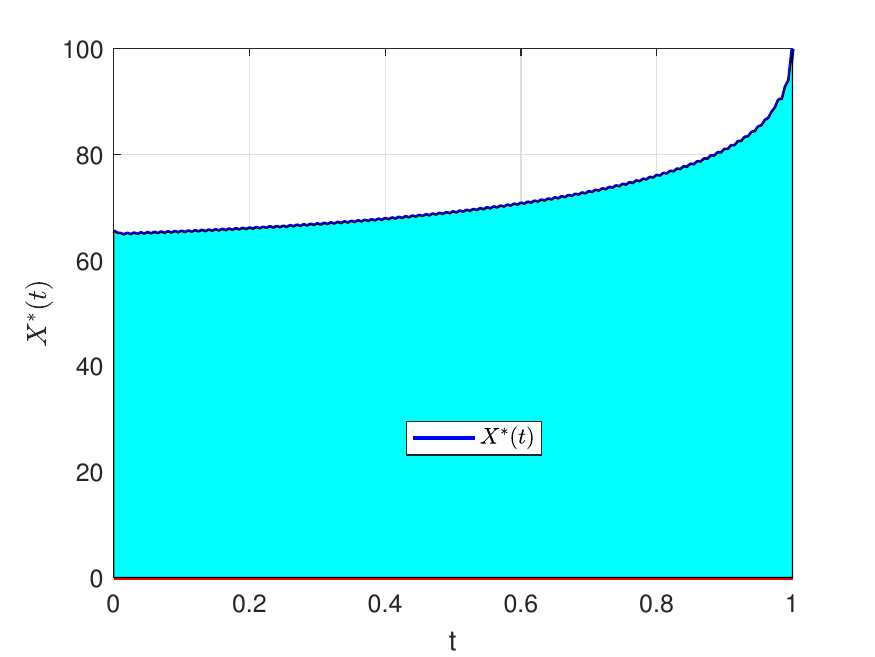}\label{Case1a}}
\hspace*{-0.3in}
\subfloat[]{ \includegraphics[width=0.52\textwidth]{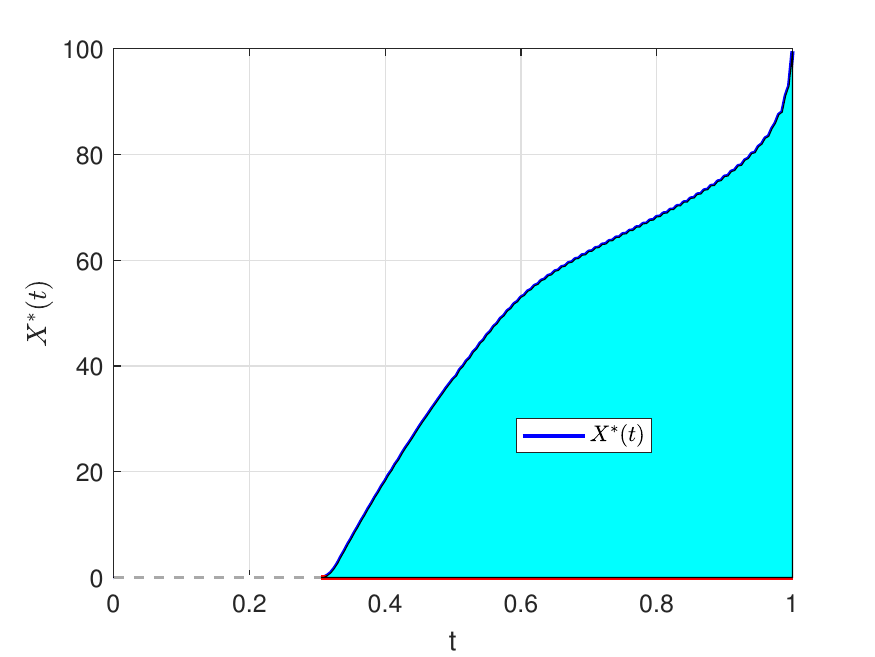}\label{Case1b}}
\caption{Single optimal exercise boundary $X^*(t)$ when a) $r(t) > 0$, b) $q(t) > 0$ but $r(t)$ changes sign in $t \in [0,T]$ from minus to plus.}
\label{F:CaseI}
\end{figure}

To determine the optimal boundary $X^*(t)$, one need to solve \eqref{Volterra} with
\begin{equation}
\Psi_1(u, X^*(u)  | t.x) = \Phi\left( d_-(x,X^*(u),t,u) \right), \qquad
\Psi_2(u, X^*(u)  | t.x) = \frac{D_q(t,u)}{D_r(t,u)} \Phi\left( d_+(x,X^*(u),t,u) \right),
\end{equation}
\noindent while taking into account the second row in Table~\ref{signChange}, which provides an explicit representation for the terminal
condition
\begin{equation}
X^*(T-) = K [\min (1,r(T)/q(T) ) ]^+.
\end{equation}
Once $X^*(t)$ is found, the American Put option price is given by a semi-analytical representation in \eqref{decompDensity2}. To illustrate the behavior of $X^*(t)$ in a test example, we use the model parameters: $K = 100, T = 1, \sigma = 0.3,  r(t) = A_r e^{-B_r t}, q(t)= A_q e^{-B_q t}, A_r = 0.05, B_r = 0.5, A_q = 0.02, B_q = 0.2$. The nonlinear equation \eqref{Volterra} can be solved in many different ways. For instance, \cite{AndersenLake2021} advocates a fixed-point iteration method that converges well since the pricing function is convex.
Alternatively, in this paper, we solve the equations using either Wolfram Mathematica's root finder combined with trapezoidal numerical integration, or Matlab's \verb+fsolve+ function with adaptive quadratures and high tolerance settings. Accordingly, the results are presented in Fig.~\ref{Case1a}.

\paragraph{Case $\bm{ q(t)> 0 \, \forall t\in[0,T]}$.} The exercise boundary's behavior changes if $r(t)$ changes sign at $t \in [0,T]$. For example, with $q(t) = 0.02$ and $r(t) = 0.03 - 0.05 e^{-1.6 t}$, this is illustrated in Fig.~\ref{Case1b}. A similar effect can be obtained by decreasing the option maturity $T$.

\subsubsection{Case $q(t) < r(t) < 0$ for all $t \in [0,T]$} \label{caseRneg}

This case has been studied in \cite{AndersenLake2021} for constant parameters satisfying $q < r < 0$. Here, we extend the analysis to time-dependent parameters $r(t)$ and $q(t)$ satisfying $q(t) < r(t) < 0, \, \forall t \in [0,T]$. Under this condition, there exist two boundaries $X^*(t)$ and $X^{**}(t)$ such that optimal exercise occurs when $X_t$ lies in the interval $[X^{**}(t), X^{*}(t)]$, see Table~\ref{signChange}. These boundaries satisfy the terminal conditions $X^*(T-) = K r(T) /q(T)$ and $X^{*}(T-) = K$.

There may exist a nonempty set of times $\mathbb{T}$ where early exercise is not optimal, and thus the exercise boundary does not exist. For $t \in \mathbb{T}$ we formally define $X^*(t)$ and $X^{**}(t)$ by assigning them some arbitrary values, such that $X^*(t) < X^{**}(t) $. This assignment does not affect the American option price since the integral in \eqref{decompDensity2} vanishes when $X^*(t) < X^{**}(t)$).

In our test example we further assume that $r(t)$ is non-decreasing and $q(t)$ is non-increasing for $t \in [0,T]$. Under these monotonicity conditions, the American Put price $P(t,x)$ is non-increasing with respect to $t \in [0,T]$ for fixed $x > 0$. Consequently, the $t$-section (a cross-section of a given area in $(t,x)$ corresponding to $t$ = constant) of $\mathcal{E}$ expands over time, manifesting as $X^{**}(t)$ being non-increasing and $X^{*}(t)$ being non-decreasing for $t \in [0,T]$.

Based on \eqref{decompDensity2} and using the same approach used to derive \eqref{Volterra}, we conclude that the pair $X^{**}(t), X^{*}(t)$ solves a system of coupled nonlinear Volterra integral equations of the second kind
\begin{align} \label{Volterra2}
K - X^*(t) &= P_E(t, X^*(t)) + \pi_2(t,X^*(t); X^*(\cdot), X^{**}(\cdot)), \\
K - X^{**}(t) &= P_E(t, X^{**}(t)) + \pi_2(t,X^{**}(t); X^*(\cdot), X^{**}(\cdot)) \nonumber \\
\pi_2(t, x; X^*(\cdot), X^{**}(\cdot)) &= \int_t^T D(t,u) r(u)K \Big[ \Phi\left(d_-(x,X^{*}(u),t,u) \right) - \Phi\left( d_-(x,X^{**}(u),t,u) \right) \Big] du \nonumber \\
&- \int_t^T D_q(t,u) q(u) x \Big[ \Phi\left( d_+(x,X^{*}(u),t,u) \right) - \Phi\left( d_+(x,X^{**}(u),t,u) \right) \Big] du, \nonumber
\end{align}
\noindent with terminal conditions $X^{**}(T-) = r(T)K/q(T), \, X^{*}(T-) = K$, and $t \in [0,T)$. Here, we use the notation $X^*(\cdot)$ to underline that $\pi_2(t, x; X^*(\cdot), X^{**}(\cdot))$ depends upon all values of $X^*(u), X^{**}(u)), \,\, \forall u \in [t,T]$.

\begin{figure}[!htb]
\vspace*{-1em}
\begin{center}
\subfloat[]{ \includegraphics[width=0.51\textwidth]{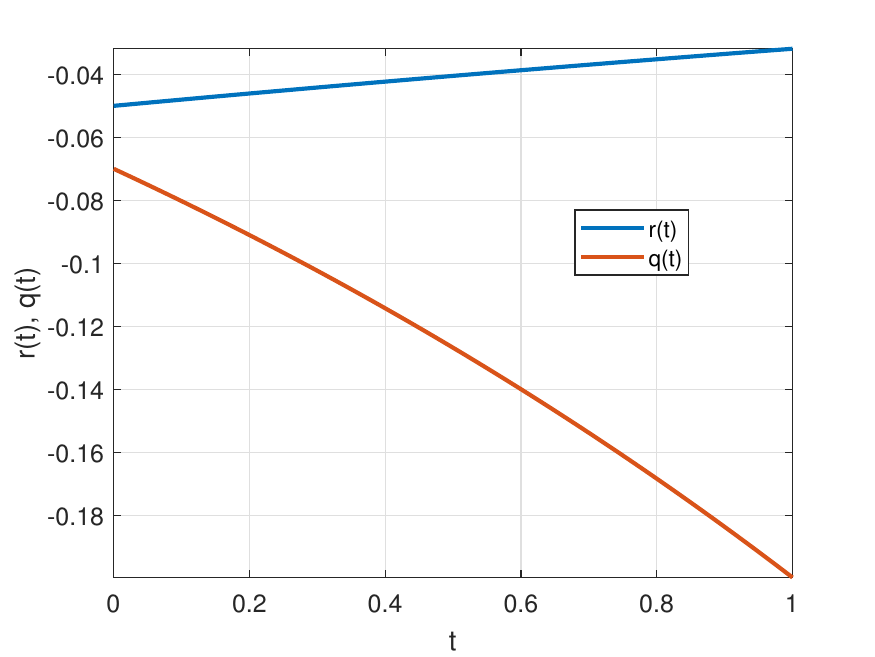} \label{case4:rq}}
\hspace*{-0.3in}
\subfloat[]{ \includegraphics[width=0.51\textwidth]{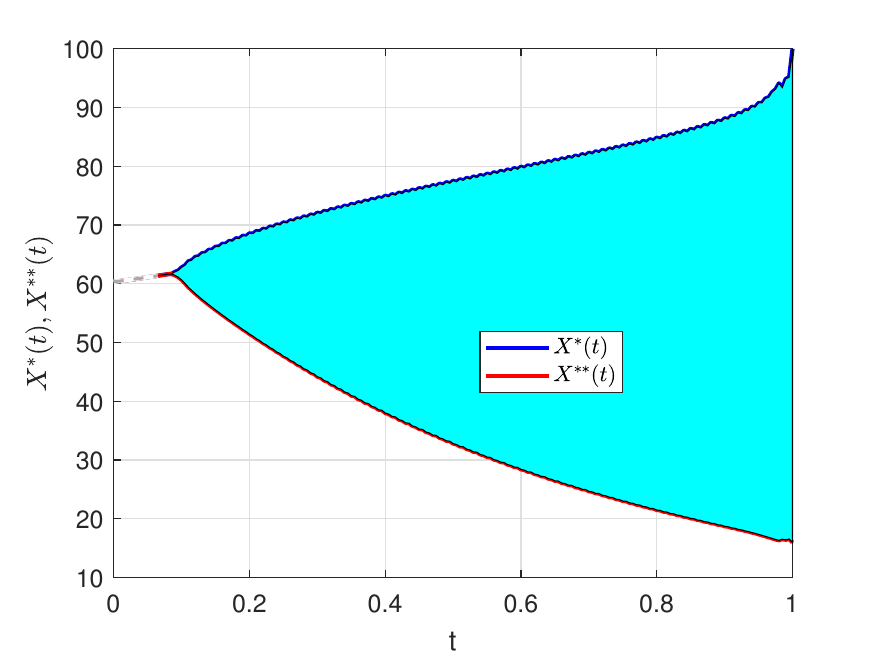} \label{case4:eb}}
\end{center}
\vspace{-1em}
\caption{Exercise regions obtained by solving \eqref{Volterra2} with parameters in Table~\ref{case4Params}: a) $r(t), q(t)$ as functions of time, b) two exercise boundaries computed in this experiment.}
\label{case4}
\end{figure}

\begin{figure}[!htb]
\vspace*{-1em}
\begin{minipage}[b]{.51\linewidth}
\captionsetup{width=0.8\linewidth}
\centering
\includegraphics[width=\linewidth]{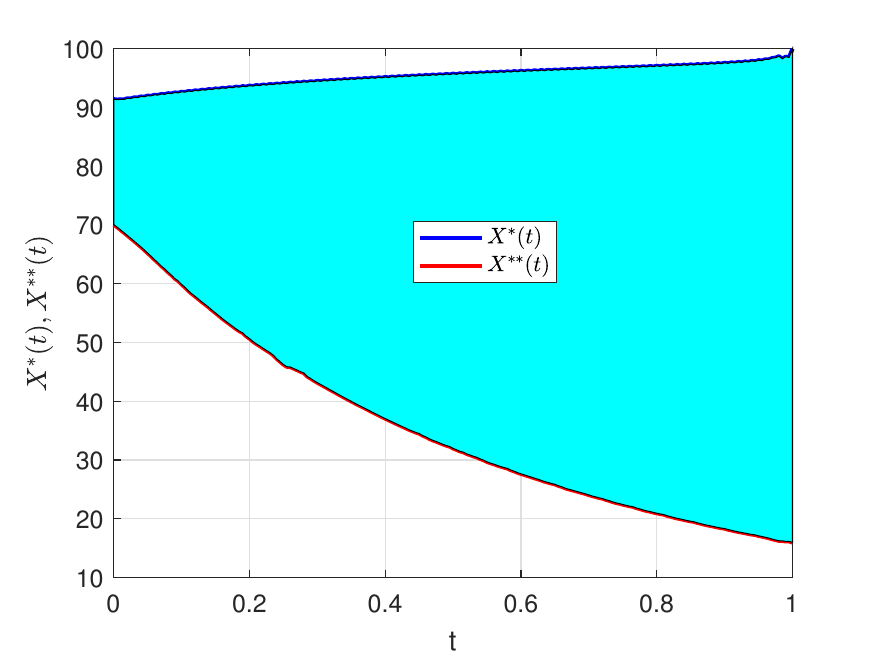}
\captionof{figure}{Exercise regions obtained by solving \eqref{Volterra2} with parameters in Table~\ref{case4Params}, but $\sigma = 0.1$.}%
\label{case4Params}%
\end{minipage}
\begin{minipage}[b]{.45\linewidth}
\captionsetup{width=0.8\linewidth}
\centering
\begin{tabular}{|r|r|r|}
\toprule
\rowcolor[rgb]{ .792,  .929,  .984}
$\bm{A_r}$ & $\bm{B_r}$ & $\bm{C_r}$ \\ \hline
-0.1  & 0.2   & 0.05 \\ \hline
$\bm{A_q}$ & $\bm{B_q}$ & $\bm{C_q}$ \\ \hline
-0.2  & -0.5  & 0.13 \\ \hline
$\bm{K}$ & $\bm{T}$ & $\bm{\sigma}$ \\ \hline
 100   & 1.0     & 0.3 \\
\bottomrule
\end{tabular}%
\captionof{table}{Parameters of the test with $q(t) < r(t) < 0$ where the exercise region contains two exercise boundaries.}
\label{case42eb}
\end{minipage}
\end{figure}

For our numerical example, we chose exponential functions for $r(t)$ and $q(t)$: $r(t) = A_r e^{-B_r t} + C_r, \, q(t) = A_q e^{-B_q t} + C_q$. The parameters used in this test are provided in Table~\ref{case4Params}, and the corresponding plots of $r(t)$ and $q(t)$ are shown in Fig.~\ref{case4:rq}. The optimal exercise boundaries $X^*(t), X^{**}(t)$ obtained by solving \eqref{Volterra2} are presented in Fig.~\ref{case4:eb}.
Thus, in this case the intersect of two boundaries occurs at $t^* > 0$. However, if we use $\sigma = 0.1$, this intersection disappears (i.e., it moves to $t^* < 0$), so the exercise boundaries look as in Fig.~\ref{case42eb}.

An analysis in \cite{ItkinKitapbayev2025} reveals that at the apparent intersection point in Fig.~\ref{case4:eb}, the boundaries don't actually intersect but rather come extremely close to each other, with the gap being less than 0.1 cents. This observation aligns with the exercise regions' structures presented in Table~\ref{signChange}.

\subsubsection{Case $q(t) < r(t) < 0$ for $t \in [0, t_1)$ while $r(t_1) = 0$ and $r(t > t_1) > 0$} \label{case21}

This case corresponds to the transition from row 3-left to row 1-left in Table~\ref{signChange}. The exercise region exhibits different characteristics before and after time $t_1$: for $t < t_1 < T$, it has two exercise boundaries, while for $T > t > t_1$, it has only one. To ensure a single regime change, we can impose additional conditions, e.g., $r'(t) \ge 0$ and $q'(t) \le 0$ for $t \in [0,T]$. Under these conditions, the interest rate is non-decreasing and transitions from negative to positive exactly once at $t_1$, while the dividend/convenience yield is non-increasing.

This case combines elements from the scenarios previously analyzed in \cref{caseRpos,caseRneg}. The exercise region's structure reflects this combination: for $t < t_1 < T$, it maintains two boundaries, while for $T > t > t_1$, it features a single non-decreasing boundary $X^*(t)$. In the latter regime, it is optimal to exercise the option when $x \le X^*(t)$. It is also possible (see \cref{caseRneg}) that for some $t \in [0,t_1)$, the $t$-section of $\mathcal{E}$ is empty.

Accordingly, the optimal boundary $X^{*}(t)$ for $t \in [t_1,T)$ is the unique solution to the nonlinear Volterra equation of the second kind
\begin{align} \label{Vol21-1}
K - X^{*}(t) &= P_E(t, X^{*}(t)) + \pi(t,X^{*}(t), X^{*}(\cdot)), \\
\pi(t, x, X^{*}(\cdot)) &= \int_t^T \Big[ D(t,u) r(u)K \Phi\left( d_-(x,X^{*}(u),t,u) \right) - D_q(t,u) q(u) x \Phi \left( d_+(x,X^{*}(u),t,u) \right) \Big] du, \nonumber
\end{align}
\noindent for $t\in[t_1,T)$ with $X^{*}(T-) = K$. In turn, the pair of optimal exercise boundaries $X^{**}(t), X^{*}(t)$ for $ t \in [0, t_1]$ solves a system of coupled nonlinear integral Volterra  equations of the second kind
\begin{align} \label{Vol21-2}
K - X^{**}(t) &= P_E(t, X^{**}(t)) + \pi_2(t,X^{**}(t);X^{**}(\cdot),X^{*}(\cdot)) \\
K - X^{*}(t) &= P_E(t, X^{*}(t)) + \pi_2(t,X^{*}(t); X^{**}(\cdot), X^{*}(\cdot)), \nonumber
\end{align}
\noindent where $\pi_2(t,X^{*}(t); X^{**}(\cdot), X^{*}(\cdot))$ is defined in \eqref{Volterra2}. It should be solved using the convention that $X^{**}(t) = 0$ for $t \in [t_1, T]$. The American Put price is then given by \eqref{decompDensitySum} which now reads
\begin{align}
P(t,x) = P_E(t,x) + \pi(t,X^{*}(t)) + \pi_2(t,x; X^{**}(t), X^{*}(t)).
\end{align}

\begin{figure}[!htb]
\vspace*{-1em}
\begin{center}
\hspace*{-0.7cm}
\subfloat[]{ \includegraphics[width=0.5\textwidth]{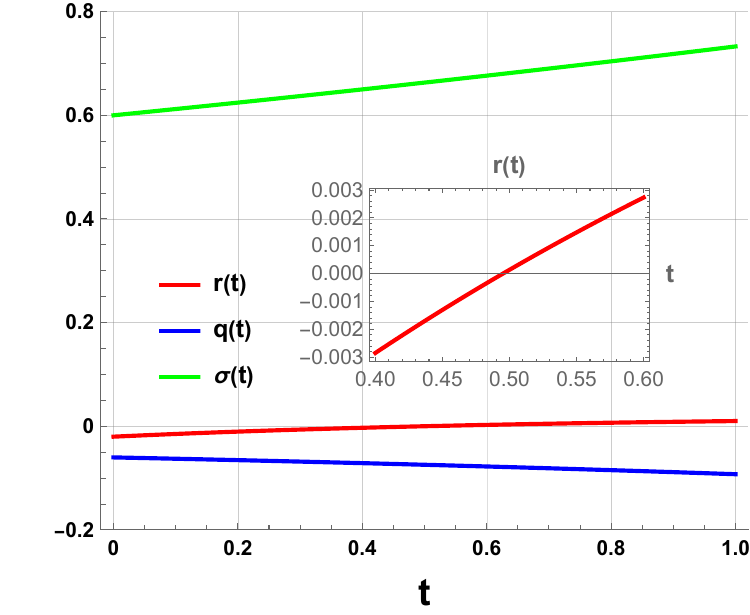} \label{case21:rqs} }
\subfloat[]{ \includegraphics[width=0.5\textwidth]{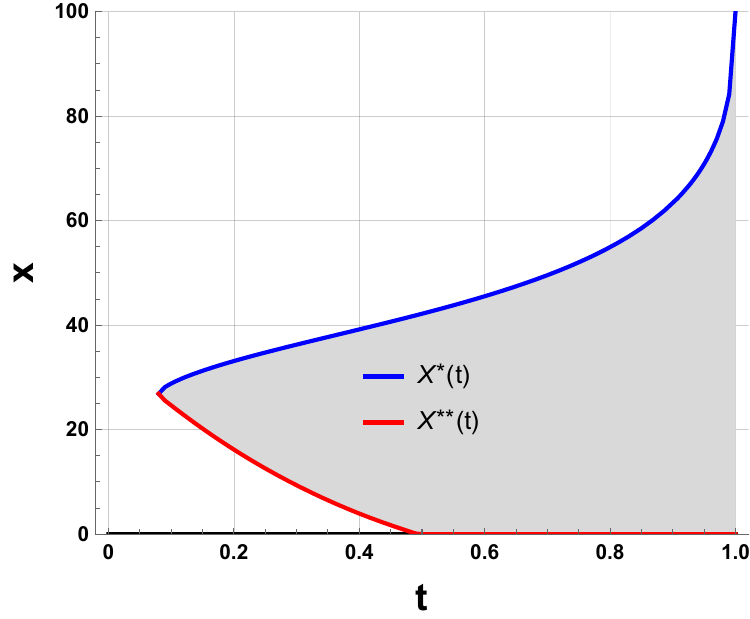} \label{case21:eb} }
\end{center}
\vspace{-1em}
\caption{Exercise regions obtained by solving \eqref{Vol21-1} together with \eqref{Vol21-2} with parameters in Table~\ref{case21Params}: a) $r(t), q(t), \sigma(t)$ as functions of time, b) the exercise boundaries.}
\label{case4}
\end{figure}

For our numerical example, again we chose exponential functions for $r(t), q(t)$ and $\sigma(t)$: $r(t) = A_r e^{-B_r t} + C_r, \, q(t) = A_q e^{-B_q t} + C_q, \, \sigma(t) = A_\sigma e^{- B_\sigma t}$. The values of parameters used in this test are provided in Table~\ref{case21Params}, and the corresponding plots of $r(t), q(t)$ and $\sigma(t)$ are shown in Fig.~\ref{case21:rqs}. The exercise boundaries $X^*(t), X^{**}(t)$ obtained by solving \eqref{Vol21-1} together with \eqref{Vol21-2} are presented in Fig.~\ref{case21:eb}.
\begin{table}[!htb]
\centering
\begin{tabular}{|r|r|r|r|r|r|r|r|r|r|}
\toprule
\rowcolor[rgb]{ .792,  .929,  .984}
$\bm{A_r}$ & $\bm{B_r}$ & $\bm{C_r}$ & $\bm{A_q}$ & $\bm{B_q}$ & $\bm{C_q}$ & $\bm{A_\sigma}$ & $\bm{B_\sigma}$ & $\bm{K}$ & $\bm{T}$ \\ \hline
-0.04  & 1.4 & 0.02 & -0.05  & -0.5  & -0.01 & 0.6 & -0.2 & 100   & 1.0 \\
\bottomrule
\end{tabular}%
\caption{Parameters of the test where $q(t)$ remains negative but $r(t)$ changes its sign at $t = t_1$.}
\label{case21Params}%
\end{table}%

\subsubsection{Case $q(t) < 0 < r(t)$ for $t \in [0, t_1)$, and $q(t) < r(t) < 0$ for $t \in (t_1, T]$} \label{complexCase}

As follows from Table~\ref{signChange}, this case is characterized by first, the existence of a single boundary which then transforms to double boundaries after $r(t)$ changes sign at $t = t_1$ and further becomes negative. As mentioned in the Remark in \cref{remark2}, this transition occurs during some interval of time that depends on $\sigma(t)$. To illustrate this behavior, for $r(t), q(t)$ we use same exponential functions as in \cref{case21}, with parameters given in Table~\ref{newCaseParams}.
\begin{table}[!htb]
\centering
\begin{tabular}{|r|r|r|r|r|r|r|r|r|r|}
\toprule
\rowcolor[rgb]{ .792,  .929,  .984}
$\bm{A_r}$ & $\bm{B_r}$ & $\bm{C_r}$ & $\bm{A_q}$ & $\bm{B_q}$ & $\bm{C_q}$ & $\bm{K}$ & $\bm{T}$ \\ \hline
0.05  & 1. & -0.03 & 0.01  & -0.8  & -0.04 & 100   & 1.0 \\
\bottomrule
\end{tabular}%
\caption{Parameters of the test where $q(t)$ remains negative but $r(t)$ changes its size at $t = t_1$ from plus to minus.}
\label{newCaseParams}%
\end{table}%
\begin{figure}[!htb]
\vspace*{-1em}
\begin{center}
\hspace*{-0.3in}
\subfloat[]{ \includegraphics[width=0.55\textwidth]{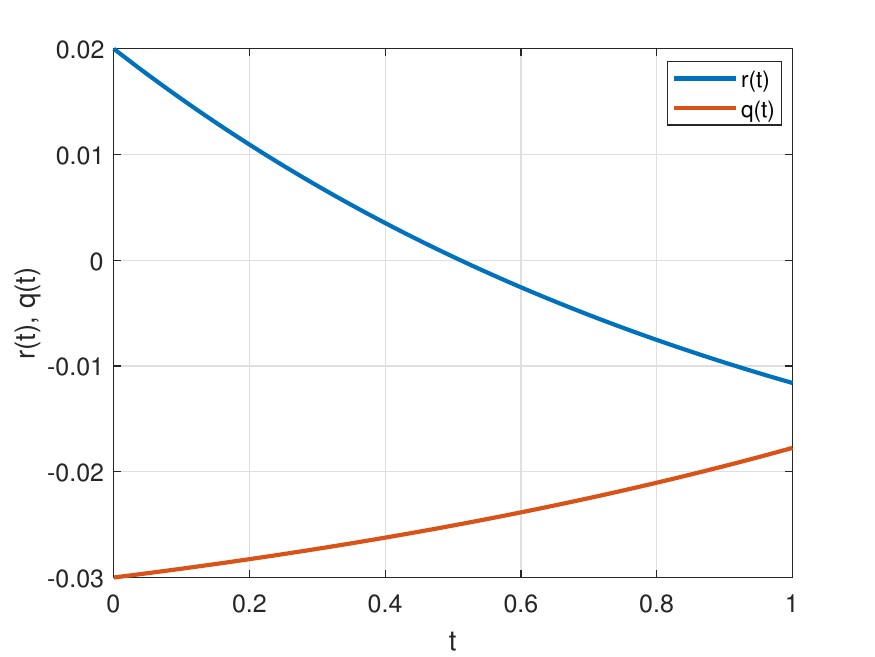} \label{newCase:rqs} }
\hspace*{-0.4in}
\subfloat[]{ \includegraphics[width=0.55\textwidth]{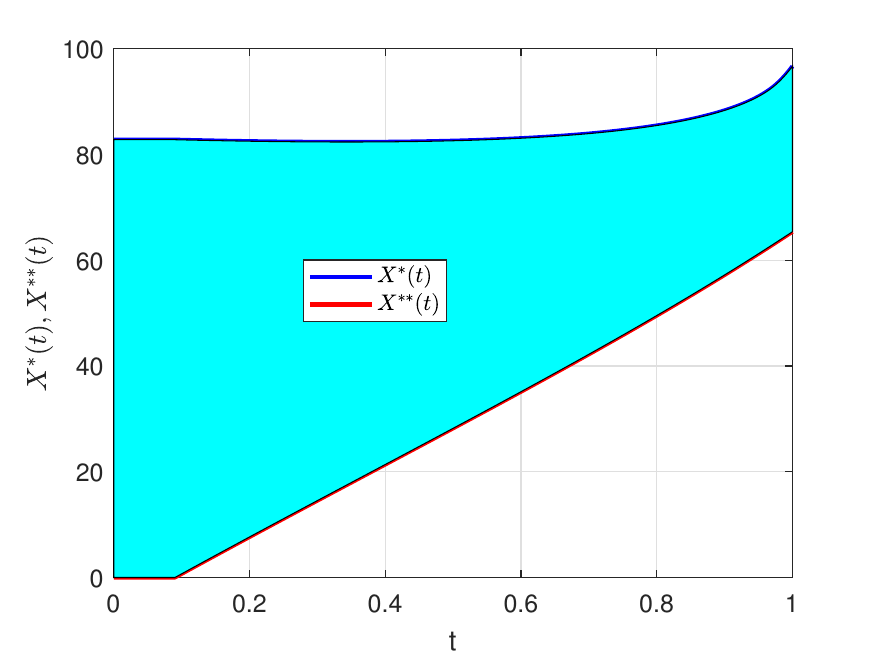} \label{newCase:eb_0_2} }
\end{center}
\vspace{-1em}
\caption{Exercise regions obtained by solving \eqref{Vol21-1} together with \eqref{Vol21-2} with parameters in Table~\ref{newCaseParams}: a) $r(t), q(t)$ as functions of time, b) exercise boundaries at $\sigma(t)$ = 0.2.}
\label{case4}
\end{figure}
\begin{figure}[!htb]
\vspace*{-1em}
\begin{center}
\hspace*{-0.3in}
\subfloat[]{ \includegraphics[width=0.55\textwidth]{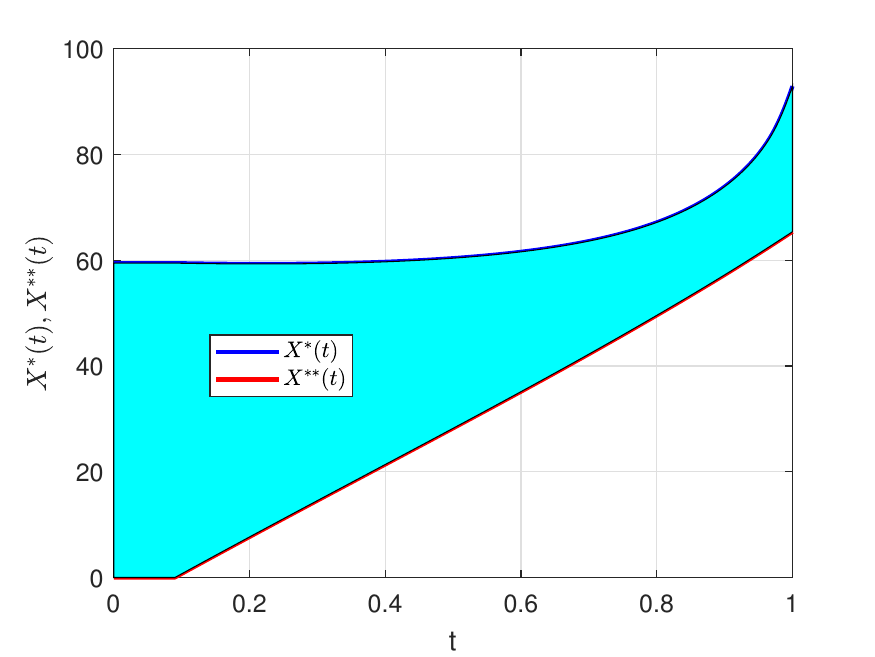} \label{newCase:eb_0_4} }
\hspace*{-0.4in}
\subfloat[]{ \includegraphics[width=0.55\textwidth]{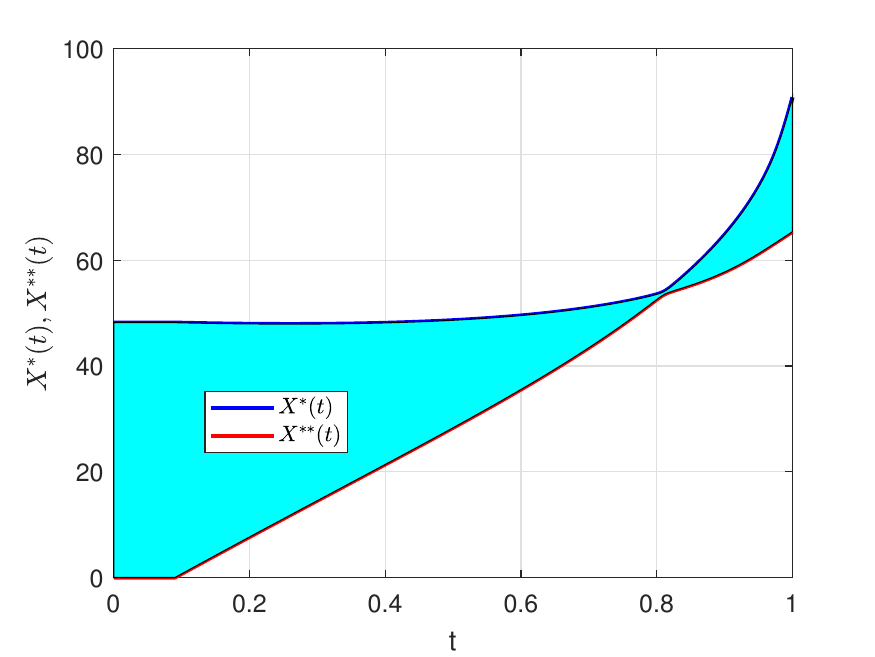} \label{newCase:eb_0_5087} }
\end{center}
\vspace{-1em}
\caption{Exercise regions obtained by solving \eqref{Vol21-1} together with \eqref{Vol21-2} with parameters in Table~\ref{newCaseParams}: a) the exercise boundaries at $\sigma(t)$ = 0.4, b) the exercise boundaries at $\sigma(t)$ = 0.5087.}
\label{case4-1}
\end{figure}

The results are shown in Fig.~\ref{case4} and \ref{case4-1}, with plots of $r(t)$ and $q(t)$ presented in Fig.~\ref{newCase:rqs}.
With $\sigma = 0.2$, Fig.~\ref{newCase:eb_0_2} reveals that the exercise region initially has a single boundary. At time $t^*  \approx 0.1 < t_1$, a second (lower) boundary emerges from zero, and both are increasing as time progresses. Similar behavior is observed for $\sigma = 0.4$ in Fig.~\ref{newCase:eb_0_4}, though here the upper and lower boundaries come closer together at $t \approx 0.8$ compared to the case shown in Fig.\ref{newCase:eb_0_2}. When $\sigma$ is further increased to 0.5087, as shown in Fig.~\ref{newCase:eb_0_5087}, the exercise boundaries intersect at $t \approx 0.82$. At this point, the original exercise region collapses, and simultaneously, a new exercise region forms with two boundaries for $t > 0.82$.

\begin{figure}[!htb]
\vspace*{-1em}
\begin{center}
\hspace*{-0.3in}
\subfloat[]{ \includegraphics[width=0.52\textwidth]{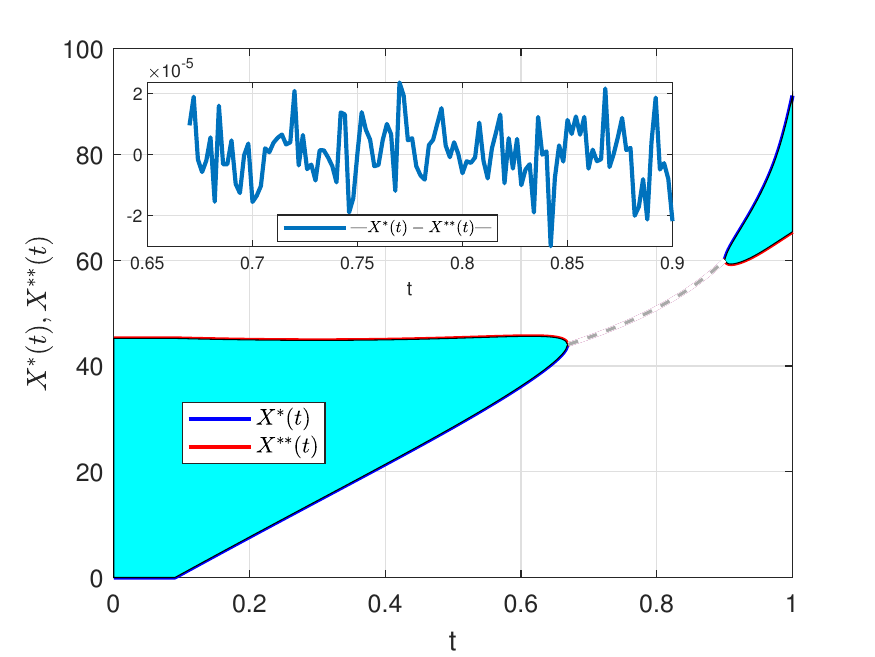} \label{newCase:eb_0_54} }
\hspace*{-0.2in}
\subfloat[]{ \includegraphics[width=0.52\textwidth]{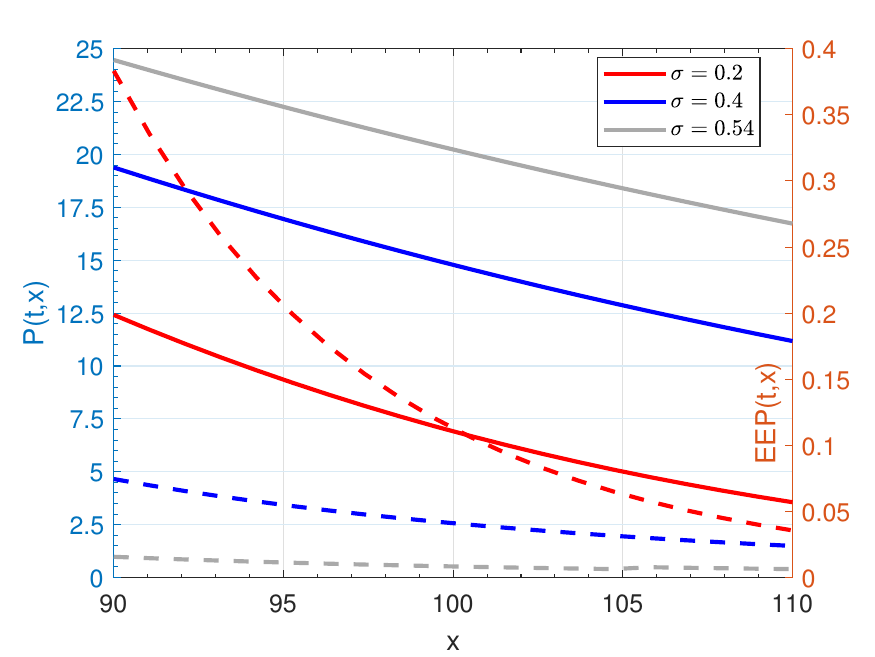} \label{newCase_Put} }
\end{center}
\vspace{-1em}
\caption{ Solution of \eqref{Vol21-1} together with \eqref{Vol21-2} with parameters in Table~\ref{newCaseParams}: a) the exercise boundaries at $\sigma(t)$ = 0.54; b) American option prices (solid lines) and EEPs (dashed lines) for various values of constant volatility $\sigma$.}
\label{putPrice}
\end{figure}

Finally, increasing the volatility to $\sigma = 0.54$ yields a more complex structure of exercise regions, as shown in Fig.~\ref{newCase:eb_0_54}.  While the exercise region initially has a single boundary, the second boundary emerges at time $t^* < t_1$. Both exercise boundaries then collapse at $t_e \approx 0.67$. A second exercise region with double boundaries appears at $t_s \approx 0.9$. For $t \in (t_e, t_s)$, the upper and lower exercise boundaries nearly coincide. By "coincide", we mean that these boundaries are very close but not exactly equal in our numerical calculations. If they were exactly equal, this would indicate the absence of an exercise region for $t \in (t_e, t_s)$.

The small difference between $X^*(t)$ and $X^{**}(t)$ could be attributed to either numerical errors (suggesting the boundaries are actually equal) or, alternatively, to our calculations being sufficiently precise to confirm the existence of a very narrow exercise region connecting the larger regions on either side. Since $r(t)$ changes its sign only once for $t \in [0,T]$, it follows from Table~\ref{signChange} that there should not be a region without exercise boundaries. To verify this, we performed an additional check confirming that the EEP remains non-negative across the entire time interval $t \in [0,T]$. This analysis not only verified the persistence of the exercise region for $t \in [t_e, t_s]$, but also uncovered an unexpected phenomenon: for $t < t_e$ the exercise boundaries switch positions, with the lower boundary becoming the upper one and vice versa, as shown in Fig.~\ref{newCase:eb_0_54}. Importantly, the EEP remains positive in both shadowed regions of the plot. On the other hand, when boundaries intersect and then diverge, it becomes ambiguous whether the previously upper boundary becomes the lower boundary or vice versa - there is no definitive way to determine which boundary is which after the crossing point.

Finally, Fig.~\ref{newCase_Put} shows the American Put option prices and their corresponding EEP, which is strictly positive. When $\sigma = 0.54$, the EEP shows monotonic behavior with respect to $x$, even though the exercise regions shown in Fig.~\ref{newCase:eb_0_54} have a complex structure.

\begin{figure}[!htb]
\vspace*{-1em}
\begin{center}
\hspace*{-0.3in}
\includegraphics[width=0.7\textwidth]{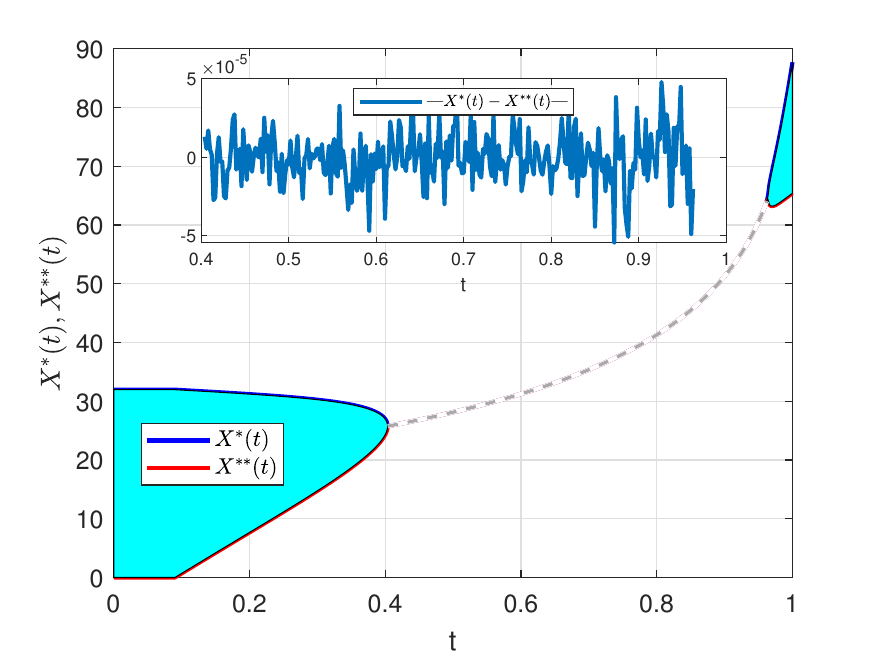}
\end{center}
\vspace{-1em}
\caption{ Solution of \eqref{Vol21-1} together with \eqref{Vol21-2} with parameters in Table~\ref{newCaseParams} and $\sigma=0.7$.}
\label{newCase:eb_0_7}
\end{figure}

If we continue to increase $\sigma$ while keeping all other parameters constant, the two islands depicted in Fig.\ref{newCase:eb_0_54} transform their shape into what is shown in Fig.\ref{newCase:eb_0_7} for $\sigma = 0.7$. Note that the gray dashed line connecting both islands is not merely a line, but rather a very narrow area (isthmus) where $X^{*}(t) > X^{**}$ in its right tail (as shown in the inset plot in Fig.~\ref{newCase:eb_0_7}), with the opposite behavior in its left tail.

It is also important to mention, that solving \eqref{Vol21-2} requires a very careful numerical procedure because otherwise it could converge to a different result due to a bad initial guess. In our calculations we used Matlab's \verb+fsolve+ function with a tolerance of $10^{-15}$ to solve this system of equations backward in time starting from $t = T$, as follows:
\begin{enumerate}
\item Solve the first equation in \eqref{Vol21-2} for $X^{*}(t)$ using the initial guess $X^{*}(t+\Delta t)$ and setting $X^{**}(t) = X^{**}(t + \Delta t)$, to get $X^{*}_1(t)$.

\item Solve the second equation in \eqref{Vol21-2} for $X^{**}(t)$ using the initial guess $X^{**}(t+\Delta t)$ and setting $X^{*}(t) = X^{*}(t + \Delta t)$, to get $X^{**}_1(t)$.

\item Solve both equations in \eqref{Vol21-2} together using the initial guess $[\alpha X^{*}_1(t), \beta X^{**}_1(t)$ to get the final solution at time $t$. Here $\alpha = \beta = 1$ if either of $X^{*}_1(t), X^{**}_1(t)$ is very small. If not, we set  $\alpha = 1.05, \beta = 0.95$ if $X^{*}_1(t) > X^{**}_1(t)$, and $\alpha = 0.95, \beta = 1.05$ otherwise.

\end{enumerate}
The first two step of this procedure are needed to determine a good initial guess for the third step.

%%%%%%%%%%%%%%%%%%%%%%%%%%%%%%%%%%%%%%%%%%%%%%%%%%%%%%%%%%%%%%%%%%%%%

\section{Models with a mean-reverting drift} \label{S:HW}

In this section, we extend the previous results using a time-inhomogeneous OU process with mean reversion. In this model the price of the underlying asset $X_t$ follows the SDE
\begin{equation} \label{HW}
dX_t = \kappa(t)[ \theta(t) - X_t] dt + \sigma(t) dW_t.
\end{equation}
Here, $\kappa(t) > 0$ represents the speed of mean-reversion, and $\theta(t)$ is the mean-reversion level. This model is widely used in Fixed Income for calibrating market rate curves (where $r_t = s(t) + X_t$, and $s(t)$ is a deterministic shift), where it is known as the Hull-White model. In commodities, it is known as the one-factor Schwartz model \citep{Schwartz1997}, applied to the logarithm of the spot price $S_t$, such that $X_t = \log(S_t)$. Since $X_t$ in this model could be negative, but we want to use it for, say cryptocurrencies, we impose an additional absorbing boundary condition $X_0 = 0$, so $(t,X_t) \in [0,T] \times [0, \infty)$.

Based on our previous analysis in \cref{S:struct}, the price of the American Put is given by \eqref{decompDensitySum} with
\begin{align} \label{HW_H}
H(u, X_u) &= r(u) \left( K - X_u \right) + \kappa(u)[ \theta(u) - X_u] = \bar{r}(u) K - \bar{q}(u) X_u, \\
\bar{q}(u) &= r(u) + \kappa(u), \qquad \bar{r}(u) = r(u) + \kappa(u) \theta(u)/K. \nonumber
\end{align}
The EEP for this model is structurally identical to the GBM case if $q(t)$ is replaced by $\bar{q}(t)$ and $r(t)$ - by $\bar{r}(t)$. However, the transition densities for the GBM model and that one in \eqref{HW} differ, necessitating an explicit representation of the Hull-White transition density to proceed.

Suppose we consider \eqref{HW} for commodities, while a more challenging case of pricing American Put on a zero-coupon bond where the instantaneous interest rate $r_t$ is stochastic and follows \eqref{HW} can be found in \cite{ItkinMuravey2024jd}. Further, assume
that $X_t$ is a spot price, e.g., a BTC price, which is influenced by supply and demand, and thus behaves like rare commodities (gold, etc.). Then, the European Put price solves the PDE, \cite{ItkinLiptonMuraveyBook} \begin{equation} \label{PDEP}
\fp{P_E}{t} + \dfrac{1}{2}\sigma^2(t) \sop{P_E}{x} + \kappa(t) [\theta(t) - x] \fp{P_E}{x} = r(t) P_E, \qquad (t, x) \in \mathbb{R}_+ \times [0, \infty),
\end{equation}
\noindent subject to the standard terminal and boundary conditions
\begin{equation} \label{tchw}
P_E(T,x) = \left(K - x\right)^+, \quad P_E(t,0) = F(0,t) K, \quad  P_E(t,x)\Big|_{x \to \infty} = 0, \quad F(0,t) = 1/D(0,t).
\end{equation}
This PDE can be reduced to the heat equation, \cite{ItkinLiptonMuraveyBook}
\begin{gather} \label{Heat}
u_\tau = u_{zz} + \lambda(t,x), \qquad \lambda(t) = \frac{K e^{-\beta(t)}}{\gamma^2(t) \sigma^2(t)}
\left[ F(0,t) \alpha (t) \left(2 k(t) \theta (t)+\alpha (t) \sigma (t)^2\right) \right], \\
u(0,z) = e^{-x} [K - x(t,z)]^+ - K F(0,T), \quad u(\tau, y(t(\tau))) = 0, \quad u(\tau, \infty) = -g(\tau), \nonumber
\end{gather}
\noindent with $(\tau , z) \in \mathbb{R}_+ \times [y(t(\tau)), \infty)$ and $\tau = \phi(t)$, $z = x \gamma(t) + y(t)$, $x(t,z) = [z - y(t)]/\gamma(t)$, and
\begin{gather} \label{tr1}
u(\tau,z) = e^{-\beta(t) - \alpha(t) x} P_E(t,x) - g(\tau), \qquad g(\tau(t)) = e^{-\beta(t)} F(0,t) K, \\
\gamma(t) = C_1 e^{\int_T^t \kappa(s) ds}, \quad \alpha(t) = C_2 e^{\int_T^t \kappa(s) ds}, \quad
y(t) = \int_t^T  \gamma(s) \left[ \kappa(s) \theta(s) + \alpha(s) \sigma^2(s) \right] ds + C_5, \nonumber \\
\phi(t) = \frac{1}{2} \int_t^T  \sigma^2(s) \gamma^2(s) ds + C_3, \quad
\beta(t) = \int_T^t \left[ r(s) - \frac{1}{2} \alpha(s)\left( 2 \kappa(s) \theta(s) + \alpha(s) \sigma^2(s)\right) \right] ds + C_4, \nonumber
\end{gather}
\noindent where $C_1,\ldots,C_5$ are some constants. In our case we can choose $C_1 = 1$, $C_2 = -1$, $C_3 = C_4 = C_5 = 0$. Accordingly, the Green's function (the transition density) of \eqref{Heat} reads, \cite{ItkinMuravey2024jd}
\begin{align}
\psi\left(z, 0 | \xi, t \right) = G(z, \xi, \tau) &= \frac{1}{2 \sqrt{\pi \tau}}\left\{\exp \left[-\frac{(z-\xi)^2}{4 \tau}\right] - \exp \left[-\frac{(z+\xi - 2 y(\tau) )^2}{4 \tau }\right]\right\}, \qquad \tau = \tau(t),
\end{align}
\noindent and the European Put option price is given by
\begin{gather} \label{U_final}
P_E(t,x) = e^{\beta(t) + \alpha(t) x} [u(\tau, z) + g(\tau(t))]\Big|_{z \to x \gamma(t) + y(t)}, \\
u(\tau, z) = \frac{1}{2\sqrt{\pi \tau}} \int_{y(0)}^{+\infty} u(0,\xi) \left[e^{-\frac{( \xi - z)^2}{4\tau}} -  e^{-\frac{(\xi + z -2 y(\tau))^2}{4\tau}} \right] d\xi  \nonumber \\
- \int_0^\tau \frac{\Psi(s,y(s)) + y'(s) g(s)}{2\sqrt{\pi (\tau - s)}} \left[e^{-\frac{(z - y(s))^2}{4(\tau-s)}} -  e^{-\frac{(z - 2 y(\tau) + y(s))^2}{4(\tau-s)}} \right] ds + \int_0^\tau \frac{g(s)}{4\sqrt{\pi(\tau - s)^3}} \Bigg[ (z- y(s)) e^{-\frac{(z-y(s))^2}{4(\tau -s)}} \nonumber \\
+ (y(s)+ z - 2 y(\tau)) e^{-\frac{(z + y(s) - 2 y(\tau))^2}{4(\tau -s)}} \Bigg] ds + \int_0^\tau \int_{y(s)}^{\infty} \frac{\lambda(s)}{2\sqrt{\pi (\tau - s)}} \left[e^{-\frac{(\xi - z)^2}{4(\tau-s)}} -  e^{-\frac{(\xi -2 y(\tau) + z)^2}{4(\tau-s)}} \right]  d\xi ds, \nonumber
\end{gather}
\noindent where the dependencies $\tau(t)$ and $z(x, t)$ are given by \eqref{tr1}. Here, the first integral and the last integral in $\xi$ can be calculated in closed form in terms of $\mathrm{Erf}$ special functions.

The function $\Psi(\tau, y(\tau))$ represents the unknown gradient of the solution at the moving boundary $y(\tau)$, defined as $\Psi(\tau, y(\tau)) =  u_z(\tau, z \rightarrow y(\tau))$. As demonstrated in \cite{ItkinMuravey2024jd}, $\Psi(\tau, y(\tau))$ solves a linear integral Volterra equation of the second kind, derived by differentiating \eqref{U_final} with respect to $x$, setting $x = y(\tau)$, and performing some regularization in limiting cases. It is important to note, that $\Psi(\tau)$ is a function of time only. Therefore this integral equation needs to be solved just once, after which computing options prices for any $x$ requires only the evaluation of integrals in \eqref{U_final}.

\subsection{Static hedge of the American option}

Looking again at \eqref{decompDensity} and the definition of $H(u, X_u)$ in \eqref{HW_H}, one can observe that \eqref{decompDensity} in this particular case can be represented in the form
\begin{align} \label{decompDensityHedge}
P \left(t, x, T \right) &= P_E \left(t, x, T \right) + \int_t^T D(t,u) \left[ \bar{r}_u P_{\CON}(u, X^*(u)) - \bar{q}
P_{\AON}(u, X^*(u)) \right] du,
\end{align}
\noindent where $P_{\CON}(u, X^*(u)), P_{\AON}(u, X^*(u))$ are Cash-or-nothing Put option and Asset-or-nothing Put options, respectively, with maturity $u$ and strike $K = X^*(u)$. Here, $X^*(t)$ can be determined independently by using the approach of \cite{ItkinMuravey2024jd}, and thus, it is treated as a known function of time. Alternatively, $P_{\CON}(u, X^*(u))$ can be treated as a barrier option with a {\it digital} payoff and the upper barrier being $X^*(t)$. Then $P_{\AON}(u, X^*(u))$ is a similar barrier {\it Call} option with zero strike.

Aside from hedging, if one needs to numerically compute the integral in \eqref{decompDensityHedge} (the EEP), a more efficient approach is to treat each integrand as the price of an Up-and-Out barrier Call option with the upper barrier: \( X^*(u) \), maturity \( u \), the payoff \( H(u, X(u)) \), and a zero strike. The pricing formula for such barrier options in the general case of the time-dependent Hull-White model was derived in \cite{ItkinLiptonMuraveyBook}. Thus, these results can be reused by substituting the relevant parameters.

Moreover, since we require a series of option prices that differ only by $X^*(u)$ and maturity $u$, these can be computed sequentially in time. By reusing prior calculations of the temporal integral up to $u - \Delta u$ and only performing new computations for the final time step $\Delta u$.

\section{Conclusion}

We provide a semi-analytical method for pricing American options by solving integral Volterra equations in time-inhomogeneous models. This approach is similar to \cite{AndersenLake2021, Peskir2007}, but extends it by covering various time-dependent models beyond the classical Black-Scholes model. We show that in such models the exercise boundaries could have "floating" structure, i.e., appear and disappear with time. Moreover, American Put and Call options in more sophisticated models from \cite{ItkinMuravey2024jd} can be priced similarly, as that paper provides explicit representations of Green's function and alternative integral Volterra equations for exercise boundaries.

%%%%%%%%%%%%%%%%%%%%%%%%%%%%%%%%%%%%%%%%%%%%%%%%%%%%%%%%%%%%%%%%%%%%%%%%%%%%%
\printbibliography[title={References}]

\end{document}